\documentclass[twocolumn,showpacs]{revtex4}
\usepackage{color}
\usepackage{graphicx}
\usepackage{dcolumn}
\usepackage{bm}

\begin{document}

\title{Strain and field modulation in bilayer graphene band structure}

\author{Hassan Raza and Edwin C. Kan}
 \address{School of Electrical and Computer Engineering, Cornell University Ithaca NY 14853 USA}%

\begin{abstract}
Using an external electric field, one can modulate the bandgap of Bernal stacked bilayer graphene by breaking $A-\tilde{B}$ symmetry. We analyze strain effects on the bilayer graphene using the extended H\"uckel theory and find that reduced interlayer distance results in higher bandgap modulation, as expected. Furthermore, above about 2.5 $\AA$ interlayer distance, the bandgap is direct, follows a convex relation to electric field and saturates to a value determined by the interlayer distance. However, below about 2.5 $\AA$, the bandgap is indirect, the trend becomes concave and a threshold electric field is observed, which also depends on the stacking distance. 
\end{abstract}


\maketitle

\begin{figure}
\vspace{2.5in}
\hskip-3.5in\includegraphics{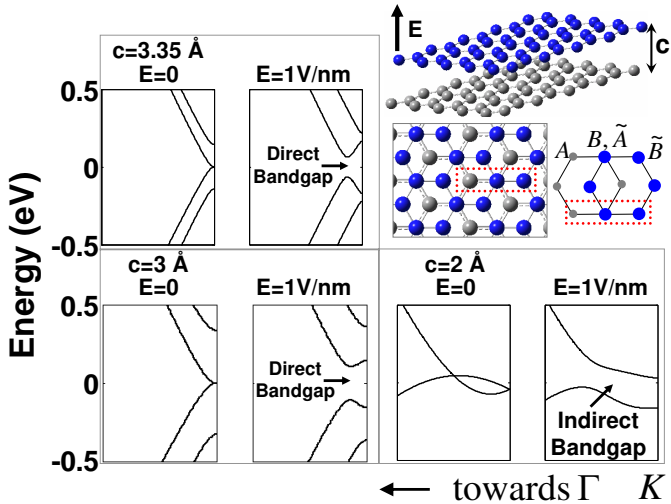}
\caption{(color online) Electronic structure of bilayer graphene. E(k) diagrams are shown around the Dirac point with and without electric fields. For equilibrium stacking distance c = 3.35 $\AA$, the dispersion is quadratic. With an external electric field, a direct bandgap appears. For c = 3 $\AA$, bandgap modulation is larger due to increased hopping between the two layers and remains direct. For c = 2$\AA$, the bilayer becomes metallic with E = 0 and has an indirect bandgap.}
\end{figure}

Graphene is a two dimensional membrane of hexagonally arranged carbon atoms. One obtains zero bandgap and linear dispersion around the Dirac point for a monolayer of graphene \cite{Wallace47, Saito98, Geim07, Neto08}. When two graphene layers are stacked in Bernal ($\tilde{A}-B$) configuration, these monolayer features are lost and the dispersion becomes quadratic \cite{McCann06_1}. The bilayer configuration is very interesting for different applications, since the bandgap can be modulated from zero to few tens of an electron volt by using an external out-of-plane electric field \cite{McCann06_1, McCann06, Min07, Castro07, Raza08, Ohto06, Oostinga07}. 

\begin{figure}
\vspace{2in}
\hskip-3.0in\includegraphics{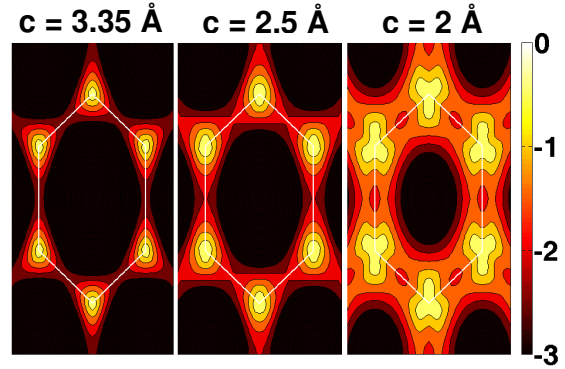}
\caption{Electronic structure of the valence band of the bilayer graphene. With decreasing stacking distance (c), new features appear in the electronic structure due to increased coupling between the two layers. The color bar is in eV.}
\end{figure}

\begin{figure}
\vspace{2in}
\hskip-3.0in\includegraphics{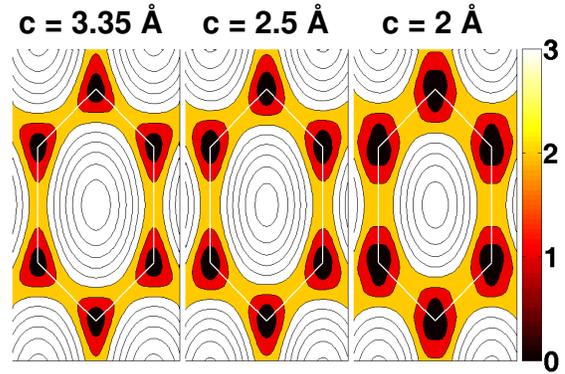}
\caption{Electronic structure of the conduction band of the bilayer graphene with various stacking distances.}
\end{figure}

This electric field breaks the $A-\tilde{B}$ symmetry in the stacked bilayers and hence lifts the degeneracy, resulting in a bandgap opening. The bandgap depends on the coupling between the two graphene layers \cite{McCann06_1, McCann06}. Therefore, one should be able to tune this bandgap by straining the bilayer. This is analogous to piezoresistivity, however the effect is enhanced exponentially due to bandgap modification. This could be very useful in sensors and other applications involving strain. Additionally, we find that by straining bilayer graphene, conduction and valence band dispersions also change. 

For electronic structure calculations, we use a semi-empirical extended H\"uckel theory (EHT) with non-orthogonal basis set. The detailed model has been described in Ref. \cite{Raza08}. The EHT parameters are transferable for different systems and have been benchmarked with the graphene band structure using generalized gradient approximation in the density function theory. This method is computationally very efficient and the non-orthogonality captures the bonding chemistry of various systems very well. It has been applied to armchair graphene nanoribbons \cite{Raza08_ac_prb}, large Si systems for electronic structure and quantum transport \cite{Raza07, Raza08_prb} and single molecule incoherent transport \cite{Raza08_prb_1, Raza08_jce} utilizing modest computational resources. 

\begin{figure}
\vspace{2.5in}
\hskip-3.25in\includegraphics{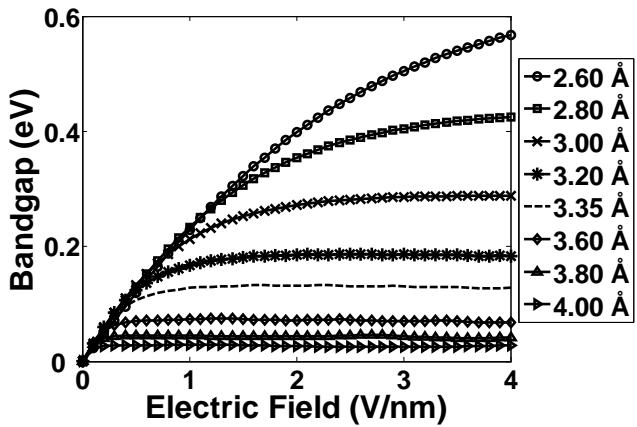}
\caption{(color online) Bandgap modulation of strained bilayer with an external electric field. The dashed line is for equilibrium c = 3.35 $\AA$, for which bandgap saturates at about 0.13 eV. A convex relation is observed as a function of electric field above c$\approx 2.5\AA$. By decreasing c, bandgap modulation increases due to increased interlayer hopping.}
\end{figure}

The C-C atomic distance is taken as 1.44 $\AA$ within the graphene plane. Out-of-plane equilibrium stacking distance (c) is 3.35 $\AA$ as shown in Fig. 1. The gray (light shaded) atoms belong to lower layer and the blue (dark shaded) atoms represent upper layer. The unit cell consists of four atoms - two atoms in each layer shown by red (black) dotted line in Fig. 1. The two gray (light shaded) atoms in the lower layer are referred to as A and B, respectively. The two blue (dark shaded) atoms in the upper layer are referred to as $\tilde{A}$ and $\tilde{B}$, respectively. The electric field is applied in the out-of-plane direction as shown in Fig. 1. Atomic visualization is done using GaussView \cite{GW03}. The substrate effects are ignored and in that sense, the graphene bilayer is assumed to be in vacuum. 

The electronic structure calculations for a bilayer graphene with equilibrium stacking distance c = 3.35 $\AA$ is shown in Fig. 1. Without any electric field, the dispersion is quadratic with a zero bandgap at the Dirac (K) point. With an electric field of 1 V/nm, a direct bandgap opens up due to $A-\tilde{B}$ symmetry breaking and the conduction/valence band minimum/maximum shifts away from the Dirac point. By decreasing c to 3 $\AA$, without an electric field, the bandgap is still zero with quadratic dispersion. However, the dispersion changes with an increase in the effective mass and the high lying bands move farther away from the Dirac point. With E = 1V/nm, again a direct bandgap is observed with the same features as the ones for c = 3.35 $\AA$. Apart from this, the bandgap modulation is higher for c = 3 $\AA$, although the Laplace's potential is smaller for the same electric field due to the reduced spacing. This is due to the increased wave function overlap between the two layers, which varies exponentially with distance as compared to the Laplace's potential which is linearly dependent on the distance. Further reducing the stacking distance results in a metallic state as shown in Fig. 1 for c = 2 $\AA$. By applying 1 V/nm electric field, an indirect bandgap is created. 

\begin{figure}
\vspace{2.5in}
\hskip-3.25in\includegraphics{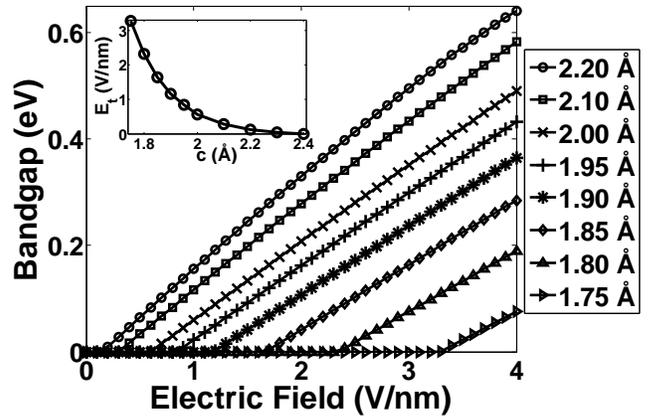}
\caption{(color online) Bandgap modulation of strained bilayer with an external electric field. Below about 2.5 $\AA$ stacking distance, a concave relation is observed as a function of electric field, where bandgap opening occurs only above a threshold electric field ($E_t$). Above $E_t$, the bandgap linearly increases with external electric field. The inset shows variation of $E_t$ with c.}
\end{figure}

Below about 2.5 $\AA$, the distance between atoms in the bottom layer and the next-nearest neighbors in the top layer becomes about 3.35 $\AA$, which is the equilibrium stacking distance for bilayer graphene. Therefore, the hopping integral between $A-\tilde{B}$ sites becomes comparable to the equilibrium one of the $\tilde{A}-B$ sites. This results in the reported electronic structure modifications in Fig. 1 for c = 2 $\AA$. This electronic structure change is also evident in the color plots of the valence band states over the two dimensional Brillouin zone in Fig. 2. With decreasing stacking distance from c = 3.35 $\AA$ to c = 2.5 $\AA$ in the absence of an electric field, the features become broad indicating an increase in effective mass. Further decreasing c to 2 $\AA$ results in directional change of the features, which signifies that the additional bonding happens in different directions than that for c = 3.35 $\AA$. These features are also present in the two dimensional Brillouin zone of the conduction band states in Fig. 3. However, it is more evident in the valence band states due to the constructive wave function overlap for the bonding nature in contrast to the anti-bonding nature of the conduction band states. 

Moreover, we find that the smaller the stacking distance, the higher the bandgap modulation by electric fields as shown in Fig. 4. This is a desirable trait for practical applications. The bandgap increases with increasing electric field and then saturates. For the equilibrium c = 3.35 $\AA$, the bandgap saturates at about 0.13 eV. For the local density approximation \cite{Min07}, the bandgap saturates at about 0.25 eV. Although the results are different for the two techniques, the general trend after straining the bilayer should remain the same in the two methods. Above about 2.5 $\AA$, the trend of field modulation is convex and in the absence of electric field, a zero bandgap is observed as shown in Fig. 4. However, below about 2.5 $\AA$, the systems becomes metallic without electric fields and the trend of field modulation becomes concave, for which a threshold behavior is observed as shown in Fig. 5. The bandgap opening occurs only above a certain threshold electric field. This threshold value monotonically increases with the decreasing stacking distance as shown in the inset of Fig. 5. As shown in Fig. 1, for c = 2 $\AA$, the valence and conduction bands are overlapping, giving rise to a metallic state. Therefore, a certain electric field needs to be applied to reduce this overlap to zero and then additional electric field opens up the bandgap. A bandgap of about 0.6 eV can be opened as shown in Figs. 3 and 4 using a 4 V/nm electric field. Calculations for such a high field are reported to show the bandgap saturation and the threshold trends for the strained bilayer. However, these high fields may not be feasible in devices due to the dielectric reliability concerns and physical constraints. High-K dielectrics may be used as an alternative to enhance field inside the graphene bilayer, while keeping the dielectric within the breakdown regime. 

We have analyzed the strain effects on the bilayer graphene using EHT and the corresponding bandgap development due to an external electric field. We find that above 2.5 $\AA$ stacking distance, the bandgap is direct and the bandgap follows a convex relation to the electric field. However, below 2.5 $\AA$ stacking distance, the bandgap becomes indirect and it follows a concave relation. 

The work is supported by National Science Foundation (NSF) and by Nanoelectronics Research Institute (NRI) through Center for Nanoscale Systems (CNS) at Cornell University. We are grateful to T. Z. Raza for GaussView \cite{GW03} visualizations and for helpful discussions.

\end{document}